\documentclass{article}

\usepackage{spconf}
\usepackage{graphicx,amsmath,amssymb}
\usepackage{fancyheadings}
\usepackage{psfrag}
\usepackage{epsfig}
\usepackage[pdftitle={DSP in Vehicles Paper Submission},pdfauthor={Author}]{hyperref}
\usepackage[ruled,vlined]{algorithm2e}

\graphicspath{ {8th DSP in Vehicles} }

\newcommand{\transp}{^{\textrm{\scriptsize{T}}}}

\begin{document}


\title{Combining Social Force Model with Model Predictive Control for Vehicle's Longitudinal Speed Regulation in Pedestrian-Dense Scenarios}

\name{Dongfang Yang$^{1}$, \"{U}mit \"{O}zg\"{u}ner$^2$}
\address{$^{1,2}$: Electrical and Computer Engineering, The Ohio State University, Columbus, OH, U.S.A. \\
         $^{1}$: E-mail: yang.3455@osu.edu}

\maketitle


\begin{abstract}
In pedestrian-dense traffic scenarios, an autonomous vehicle may have to safely drive through a crowd of pedestrians while the vehicle tries to keep the desired speed as much as possible. This requires a model that can predict the motion of crowd pedestrians and a method for the vehicle to predictively adjust its speed. In this study, the model-based predictive control (MPC) was combined with a social-force based vehicle-crowd interaction (VCI) model to regulate the longitudinal speed of the autonomous vehicle. The predictive feature of the VCI model can be precisely utilized by the MPC. A criterion for simultaneously guaranteeing pedestrian safety and keeping the desired speed was designed, and consequently, the MPC was formulated as a standard quadratic programming (QP) problem, which can be easily solved by standard QP toolbox. The proposed approach was compared with the traditional proportional-integral-derivative (PID) control approach for regulating longitudinal speed. Scenarios of different pedestrian density were evaluated in simulation. The results demonstrated the merits of the proposed method to address this type of problem. It also shows the potential of extending the method to address more complex vehicle-pedestrian interaction situations.

\end{abstract}



\section{Introduction}

Pedestrian safety has always been the main concern of the traffic system. In U.S., from year 2007 to 2016, the percentage of pedestrian fatalities in total fatalities has increased from 11\% to 16\%\cite{NHTSA16}. In the statistics of 2016, 72\% of pedestrian fatalities does not happen at intersection, which means these fatalities happen at places where there is no traffic signal controlling the priorities of the traffic participants. Therefore, it is important to study such unsignalized scenarios, especially for autonomous vehicles that have the ability to automatically adjust their speed hence avoid errors caused by human drivers. In this study, a specific unsignalized scenario is considered, in which the autonomous vehicle needs to drive through a crowd of pedestrians. Pedestrian crowd with high density (e.g. larger than 10 pedestrians) is the main focus. 

To cope with this scenario, this study proposed a model-based predictive control strategy that incorporates a pedestrian motion prediction model to achieve the longitudinal speed regulation. Model predictive control (MPC) \cite{camacho2013model} has been used for vehicle longitudinal speed regulation for a long time. Most studies about MPC longitudinal speed regulation focuses on problems such as keeping a desired distance to the front vehicle, e.g., adaptive cruise control \cite{bageshwar2004model}, or dealing with big obstacles that appear in the front, e.g., yielding to a cut-in vehicle \cite{liu2015predictive}. This study entered into a new area of using MPC for vehicle-pedestrian interaction scenario, which has not been properly addressed yet. 

Since pedestrian motion is affected by various factors, it is necessary to find an appropriate model that can effectively describe the vehicle-pedestrian interaction. Most existing studies explored only the interaction between the vehicle and a couple of pedestrians, in which theories such as gap acceptance and time to collision have been used to determine the pedestrian's intention~\cite{chen2017evaluation}. In~\cite{chao2015vehicle}, a social force model~\cite{helbing1995social} appended with the vehicle influence was proposed to predict the pedestrian's motion for the vehicle speed regulation. However, this model doesn't consider the interaction with high-density pedestrians and the proposed vehicle speed regulation method has no predictive ability. In our previous work \cite{yang2018social}, a social force based vehicle-crowd interaction (VCI) model was proposed to predict the motion of pedestrian crowd of any density, in which each individual pedestrian motion is subject to surrounding pedestrians and incoming vehicles. This study combined our previous work with the long-established MPC by designing the customized state constraints and the cost function to address the longitudinal speed regulation problem. The proposed method can keep the vehicle from a safe distance to the closest pedestrian in front, and in the meantime, try to maintain its desired speed as much as possible.

\begin{figure}
	\centering
	\includegraphics[width=\linewidth]{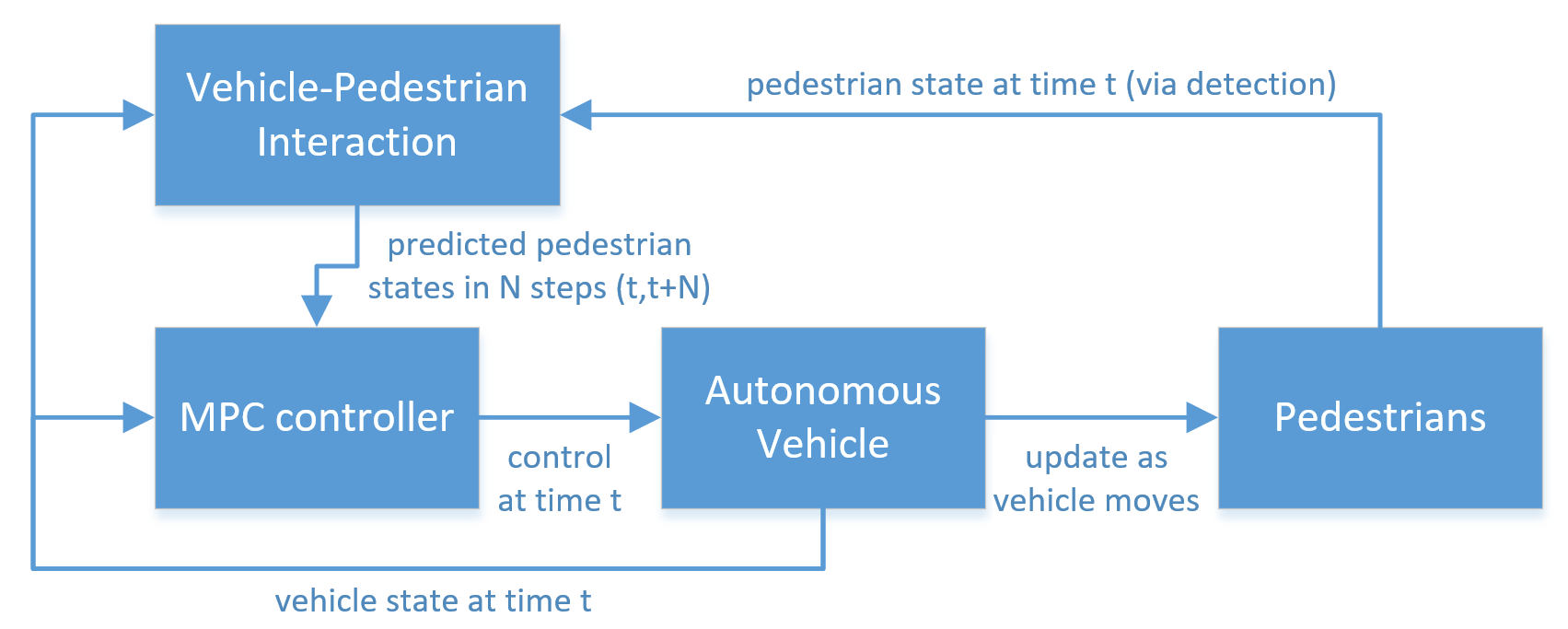}
	\caption{The structure of the proposed MPC-VCI longitudinal speed regulation strategy }
	\label{fig:structure}
\end{figure}

The flowchart of the combined MPC-VCI method is illustrated in figure~\ref{fig:structure}. Vehicle-pedestrian interaction is evaluated at each time step $k$. The future $N$-step pedestrian motion is then predicted and provided with MPC. MPC utilizes the predicted motion to formulate a standard quadratic programming (QP) problem. By solving this QP problem, the optimized control is obtained and consequently applied to the vehicle.  

The rest of the paper is organized as follows. Section 2 outlines the vehicle-crowd interaction model with emphasis of the slightly modified vehicle influence. Section 3 details the MPC based longitudinal speed regulation policy, which includes vehicle dynamics, MPC synthesis, QP generation, and complete algorithm. Section 4 presents the evaluation procedure, followed by simulation results in section 5. In the end, section 6 concludes this study and discusses future work.


\section{Pedestrian Motion Prediction}
\subsection{Vehicle-Crowd Interaction Model}
A social force based vehicle-crowd interaction (VCI) model~\cite{yang2018social} is used for the pedestrian motion prediction under the vehicle influence. In this model, each pedestrian motion $x_i\in\mathbb{R}^2$ is governed by 2D planar point-mass Newtonian dynamics subject to a total force $F_i\in\mathbb{R}^2$ consisting of several sub-forces:
\begin{equation}
	\frac{d^2x_i}{dt^2}=\frac{v_i}{dt}=a_i=\frac{F_i}{m},
	\label{eq:newton_dynamics}
\end{equation}
where
\begin{equation}
    \notag
    F_i=\sum_{j\in \mathbb{Q}(i)}(f_r^{ij}+f_c^{ij}+f_n^{ij})+f_v^i+\beta_i(f_v^i) \cdot f_d^i.
\end{equation}
$j\in\mathbb{Q}(i)$ denotes the index of nearby pedestrians around pedestrian $i$. $f_r^{ij}, f_c^{ij}, f_n^{ij}$ are social forces, which denote the repulsive (attractive) force, the collision force, and the navigational force from pedestrian $j$ to pedestrian $i$, respectively. $f_v^i$ is the vehicle influence on pedestrian $i$, which is also a repulsive force with a specific direction. $f_d^i$ is the destination force that drives the pedestrian to the temporary destination. $\beta_i(f_v^i)\in[0,1]$ is a scalar that adjusts the magnitude of $f_d^i$. Details of modeling the above sub-forces can be found in~\cite{yang2018social}. 

\subsection{Vehicle Influence}

\begin{figure}
	\centering
	\includegraphics[width=\linewidth]{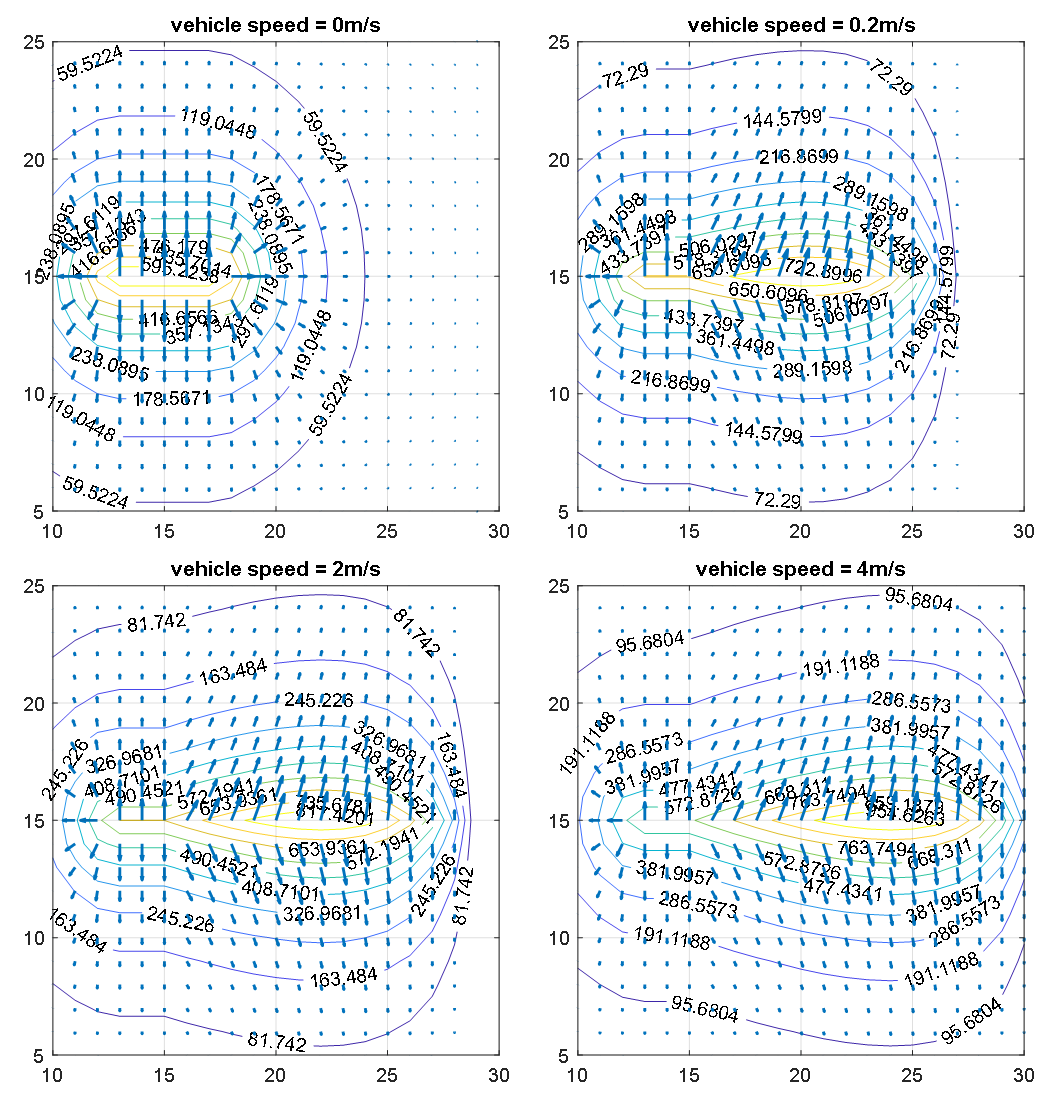}
	\caption{The contour plot of vehicle influence at different longitudinal speed. The vehicle is located at $(15,15)\transp$, facing positive x-axis. The vehicle length is 5, and vehicle width is 2. Blue arrows indicate the direction and the magnitude (arrow length) of vehicle influence force on a pedestrian located at the arrow position. As the longitudinal speed increases, the influence area expands.}
	\label{fig:veh_influence}
\end{figure}

Based on the original vehicle influence proposed in~\cite{yang2018social}, some parameters of the model was slightly modified to be more suitable for this specific interaction problem. When the vehicle longitudinal speed is less than $0.2m/s$, the vehicle is simply regarded as a rectangular static obstacle. Roughly speaking, the vehicle influence can be viewed as a potential field subject to the change of the vehicle speed. Figure~\ref{fig:veh_influence} shows the magnitude and the direction of $f_v^i$ in the surrounding area of the vehicle with different vehicle longitudinal speed. 

\subsection{Motion Prediction}

To predict pedestrian motion, it's assumed that all pedestrian state at time $t=t'$ can be correctly obtained, which means the sensing capability of the vehicle is perfect. In the prediction process, it is also assumed that the vehicle moves at a constant longitudinal speed the same as the current speed. Pedestrian motion at $t>t'$ is then calculated by iteratively applying the pedestrian dynamics in equation~\eqref{eq:newton_dynamics}.

\section{Longitudinal Speed Regulation}
\subsection{Vehicle Dynamics}
This study only considers longitudinal speed regulation, so a planner vehicle model with only longitudinal dynamics~\cite{liu2015predictive} is sufficient for this study:
\begin{equation}
    M\Ddot{s}(t)+\alpha\Dot{s}(t)=F_t(t)-F_b(t)
    \label{eq:longi_dyn}
\end{equation}
where $s$ is vehicle longitudinal position, $M$ is the vehicle mass, $\alpha$ is a linearized friction coefficient, and $F_t, F_b$ are traction force and brake force of the vehicle, respectively.

Let $x=[x_1,x_2]\transp=[s,\Dot{s}]\transp\in \mathbb{R}^2$ be a state vector of the vehicle position and speed. Equation~\eqref{eq:longi_dyn} can be written as matrix form. Furthermore, with discretization time $\Delta t$, the discretized vehicle dynamics can be obtained:
\begin{equation}
    x(k+1) = Ax(k)+Bu(k)
    \label{eq:discrete_dynamics}
\end{equation}
where
\begin{equation}
    \notag
    A = 
    \begin{bmatrix}
        1 & \Delta t \\
        0 & 1-\frac{\alpha\Delta t}{M}
    \end{bmatrix},
    B = 
        \begin{bmatrix}
        0 \\
        \frac{\Delta t}{M}
    \end{bmatrix},
    u(k) = F_t(k)-F_b(k).
\end{equation}

\subsection{Model Predictive Controller (MPC) Synthesis}

At time step $k$, with the vehicle dynamics~\eqref{eq:discrete_dynamics} and current vehicle state $x(k)$, future vehicle state $x(k+n)$ can be obtained by iteratively applying the vehicle dynamics:
\begin{multline}
    x(k+n|k) = A^nx(k)+A^{n-1}Bu(k|k)+A^{n-2}Bu(k+1|k)+\\
    \dots+ABu(k+n-2|k)+Bu(k+n-1|k).
\end{multline}
For simplicity, $x(k+n)$ will be used instead of $x(k+n|k)$ for the rest of the paper.

Now, consider a MPC with $N$-step prediction horizon. The vehicle state from steps $k+1$ to $k+N$ can be combined and represented as following equation:
\begin{equation}
    X=S_xx_k+S_uU
    \label{eq:mpc_synthesis}
\end{equation}
where
\begin{equation}
    \notag
    X=
    \begin{bmatrix}
        x(k+1)\\
        x(k+2)\\
        \vdots \\
        x(k+N)
    \end{bmatrix}\in\mathbb{R}^{2N},
    S_x=
        \begin{bmatrix}
        A\\
        A^2\\
        \vdots \\
        A^N
    \end{bmatrix}\in \mathbb{R}^{2N\times 2},
\end{equation}
\begin{equation}
    \notag
    S_u=
    \begin{bmatrix}
        B & 0 & \dots & 0 \\
        AB & B & \dots & 0\\
        \vdots & \ddots & \ddots & \vdots \\
        A^{N-1}B & \dots & AB & B
    \end{bmatrix}\in \mathbb{R}^{2N\times N}, 
\end{equation}
\begin{equation}
    \notag
    U=
    \begin{bmatrix}
        u(k)\\
        u(k+1)\\
        \vdots \\
        u(k+N-1)
    \end{bmatrix}\in\mathbb{R}^{N},x_k=x(k)\in\mathbb{R}^2.
\end{equation}

Due to the physical limitation of the vehicle, there are constraints on both the control action and the control action rate. Hence, $\forall i = k+1, \dots, k+N$, we have
\begin{align}
    |u(i)| &\le u_{max} \\
    |\Delta u(i)| &\le \Delta u_{max}.
\end{align}
A speed constraint is also considered:
\begin{equation}
    v_{min}\le x_2(i) \le v_{max}, \forall i = k+1, \dots, k+N.
\end{equation}

To avoid collision between the vehicle and pedestrians, a safe distance $d_{safe}$ to the closest pedestrian in front of the vehicle must be maintained all the time. Since the previous defined pedestrian model can predict the pedestrian motion under vehicle influence in future N steps, the predicted pedestrian information can be used by MPC to maintain the safe distance in N steps. 

This pedestrian safety is formulated as a hard constraint on the vehicle state:
\begin{equation}
    x_1(i)-x_p(i)\ge d_{safe}, \forall i = k+1, \dots, k+N
\end{equation}
where $x_p(i)$ is position of the closest pedestrian in front of the vehicle in the axis of vehicle's longitudinal position.

\subsection{Cost Function Design and Quadratic Programming (QP) Problem Formulation}
The ultimate goal of the MPC is to find a control sequence $U=[u(k),u(k+1),\dots,u(k+N-1)]\transp$ at every $x(k)$ such that the vehicle obeys both the state constraints and the pedestrian safety requirement, while in the meantime, tries to keep the desired speed $v_d$ as much as possible. Therefore, the cost function is designed as how close the vehicle will keep its desired speed:
\begin{equation}
    J(k)=(A_rX-V_r)\transp Q(A_rX-V_r)    
\end{equation}
where $Q$ is the quadratic cost, $V_r=[v_r,v_r,\dots,v_r]\transp\in\mathbb{R}^{N}$ represents the reference speed in N steps, and
\begin{equation}
    \notag
    A_r=
    \begin{bmatrix}
        0 & 1 & 0 & 0 & \dots & 0 & 0 \\
        0 & 0 & 0 & 1 & \dots & 0 & 0 \\
        \vdots & \vdots & \vdots & \vdots & \ddots & \vdots & \vdots \\
        0 & 0 & 0 & 0 & \dots & 0 & 1 
    \end{bmatrix}\in\mathbb{R}^{N\times 2N}
\end{equation}
which extracts the velocity states from $X$.

Substituting equation~\eqref{eq:mpc_synthesis}, we can rewrite the cost function as:
\begin{equation}
    J(k)=U\transp HU+2FU+Y
\end{equation}
where 
\begin{align}\notag
    H&=S_u\transp A_r\transp QA_rS_u \\ \notag
    F&=(A_rS_xx_k-V_r)\transp QA_rS_u \\ \notag
    Y&=(A_rS_xx_k-V_r)\transp Q(A_rS_xx_k-V_r)=const.
\end{align}
Similarly, by substituting equation~\eqref{eq:mpc_synthesis} and rearrange the constraint equations, the state constraints can be rewritten as:
\begin{align}
    \label{eq:c1} A_uU &\ge -U_{max} \\
    \label{eq:c2} -M_uU &\ge -\Delta U_{max}-u_0 \\
    \label{eq:c3} M_uU &\ge -\Delta U_{max}+u_0 \\
    \label{eq:c4} -M_vS_uU &\ge -V_{max}+M_vS_xx_k \\
    \label{eq:c5} M_vS_uU &\ge V_{max}-M_vS_xx_k \\
    \label{eq:c6} -M_xS_uU &\ge D_{safe}-X_{p}+M_xS_xx_k
\end{align}
where
\begin{align}
    \notag U_{max}&=[u_{max},u_{max},\dots,u_{max}]\transp\in\mathbb{R}^{2N},\\
    \notag \Delta U_{max}&=[\Delta u_{max},\Delta u_{max},\dots,\Delta u_{max}]\transp\in\mathbb{R}^{N},\\
    \notag u_0&=[u(k-1),0,\dots,0]\transp\in\mathbb{R}^N,\\ 
    \notag V_{max}&=[v_{max},v_{max},\dots,v_{max}]\transp\in\mathbb{R}^{N},\\
    \notag V_{min}&=[v_{min},v_{min},\dots,v_{min}]\transp\in\mathbb{R}^{N},\\
    \notag D_{safe}&=[d_{safe},d_{safe},\dots,d_{safe}]\transp\in\mathbb{R}^{N},\\
    \notag X_{p}&=[x_{p}(k+1),x_{p}(k+2),\dots,x_{p}(k+N-1)]\transp\in\mathbb{R}^{N},
\end{align}
\begin{equation}\notag
    A_u=
    \begin{bmatrix}
        -1& 0&\dots&0\\
         1& 0&\dots&0\\
         0&-1&\dots&0\\
         0& 1&\dots&0\\
        \vdots&\vdots&\ddots&\vdots\\
         0& 0&\dots&-1\\
         0& 0&\dots&1
    \end{bmatrix}\in\mathbb{R}^{2N\times N},
\end{equation}
\begin{equation}\notag
    M_u=
    \begin{bmatrix}
         1& 0&\dots&0&0\\
        -1& 1&\dots&0&0\\
         0&-1&\dots&0&0\\
        \vdots&\vdots&\ddots&\vdots&\vdots\\
         0& 0&\dots&-1&1\\
         0& 0&\dots&0&-1
    \end{bmatrix}\in\mathbb{R}^{N\times N},
\end{equation}
\begin{equation}\notag
    M_v=A_r,M_x=
    \begin{bmatrix}
        1 & 0 & 0 & 0 & \dots & 0 & 0 \\
        0 & 0 & 1 & 0 & \dots & 0 & 0 \\
        \vdots & \vdots & \vdots & \vdots & \ddots & \vdots & \vdots \\
        0 & 0 & 0 & 0 & \dots & 1 & 0 
    \end{bmatrix}\in\mathbb{R}^{N\times 2N}.
\end{equation}

Finally, the optimal control sequence $U^*$ can be obtained by solving the following standard QP problem:
\begin{equation}
    \label{eq:qp}
    U^*=\underset{U}{\arg\min}(U\transp HU+2FU+Y)
\end{equation}
subject to equations from~\eqref{eq:c1} to~\eqref{eq:c6}.

This QP problem can be easily solved by standard QP toolbox. In this study, the 'mpcqpsolver' in Matlab is used to solve this QP problem. Once $U^*$ is obtained, The first action of $U^*$ is applied for the current step.  

\subsection{MPC Feasibility and Supplementary Proportional-integral-derivative (PID) Control}

\begin{figure}
	\centering
	\includegraphics[width=0.9\linewidth]{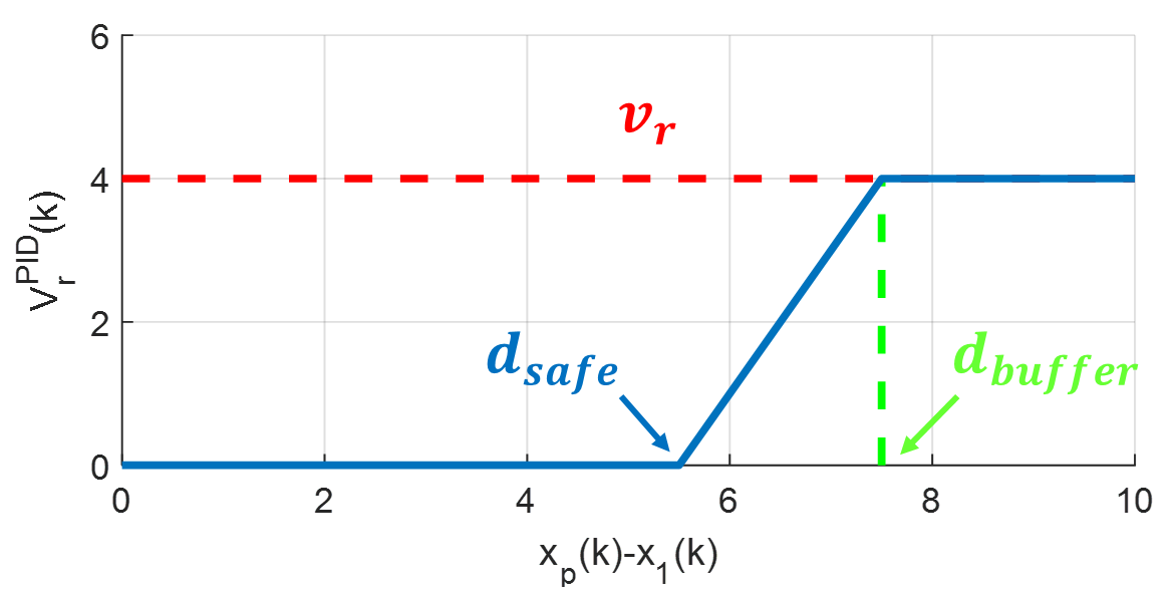}
	\caption{The longitudinal reference speed for the PID controller. A buffer distance is designed to gradually change the reference speed.}
	\label{fig:pid_ref_speed}
\end{figure}

Due to the highly uncertainty of the pedestrian-dense traffic scenario, solving the above QP problem for MPC might be infeasible. In this study, a classical PID approach~\cite{franklin1998digital} is proposed to supplement the MPC. At any time step, when the MPC cannot find a feasible solution, the controller switches to PID approach. To maintain the safe distance $d_{safe}$, a reference longitudinal speed $v_r^{PID}(k)$ for PID is determined based on the current distance to the closest pedestrian in front $x_p(k)-x_1(k)$, as shown in figure~\ref{fig:pid_ref_speed}. The discrete-time PID control action is obtained as follows:
\begin{equation}
    u(k)=-[u_p(k)+u_i(k)+u_d(k)]
\end{equation}
where
\begin{align}
    \notag u_p(k)&=K_pe(k),\\
    \notag u_i(k)&=K_ie(k)\Delta t+u_i(k-1),\\
    \notag u_d(k)&=K_d[e(k)-e(k-1)]/\Delta t,\\
    \notag e(k)&=x_2(k)-v_r^{PID}(k),
\end{align}
and $K_p, K_i, K_d$ are PID parameters.

\begin{table*}[]
	\centering
	\caption{Parameters for the longitudinal speed regulation}
	\label{tab:para_tuned}
	\begin{tabular}{||c|c|c||c|c|c||c|c|c||}
		\hline
		Parameter&Tuned Value&Unit &Parameter&Tuned Value&Unit&Parameter&Tuned Value&Unit \\ \hline
		$M$ & 1000 &  kg         & $\alpha$  &100 & $N(m/s)^{-1}$ &$\Delta t$& 0.05 & s \\ \hline
        $u_{max}$ & 8000 &  N    & $v_r$     & 4 & m/s            & $N$ & 15 &   \\ \hline
        $\Delta u_{max}$&1000&N  &$K_p$     & 300 &               & $d_{safe}$& 8 & m \\ \hline
        $v_{max}$ & 20 & m/s     & $K_i$     & 10 &               & $d_{buffer}$& 10 & m \\ \hline
        $v_{min}$ & 0 & m/s      & $K_d$     & 100 &              & $Q$     & $I_N$ & \\ \hline
                    
	\end{tabular}
\end{table*}

\subsection{Overall Algorithm}
Table~\ref{tab:para_tuned} shows all the parameters for the vehicle dynamics, the MPC, and the PID, which are manually tuned in the simulation. The overall algorithm to regulate the longitudinal speed of the autonomous vehicle is summarized in algorithm~\ref{al:mpc}.
\begin{algorithm}
    \label{al:mpc}
    \SetAlgoLined
    \caption{MPC+PID longitudinal speed regulation}
    \KwResult{control action $u(k)$}
        initialization\;
        \For{each time step $k$}{
            obtain $x(k)$ and all pedestrian states\;
            predict pedestrian motion and obtain $X_p$\;
            solve for $U^*=\underset{U}{\arg\min}(U\transp HU+2FU+Y)$\;
            \eIf{MPC is feasible}{
            apply $u(k)=U^*(1)$\;
            }{
            apply $u(k)=f_{PID}(x(k),x_p(k))$\;
            }
        }
    
\end{algorithm}


\section{Evaluation}
\begin{figure}
	\centering
	\includegraphics[width=0.8\linewidth]{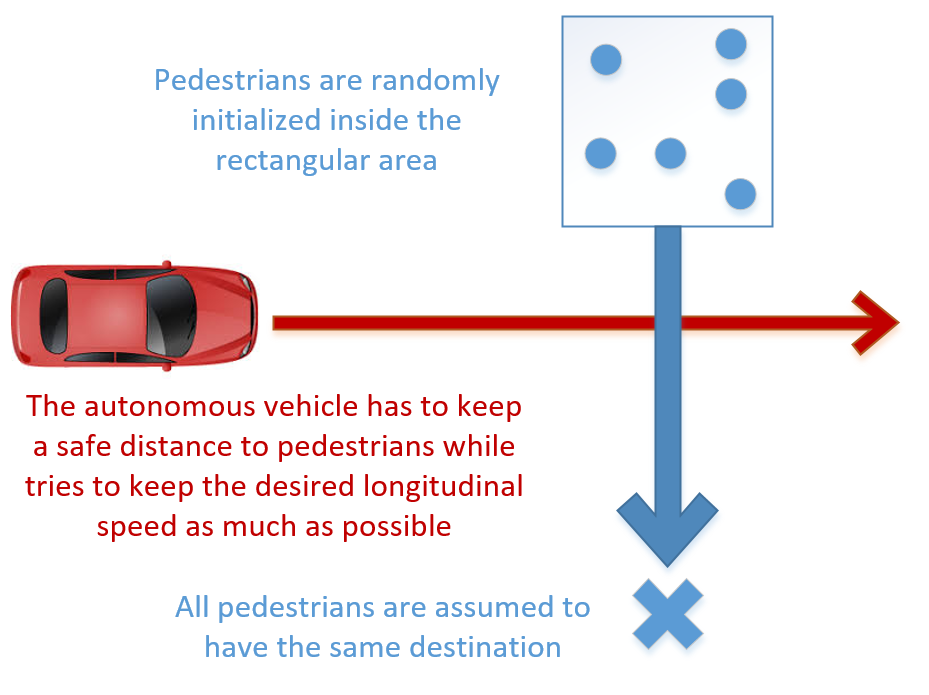}
	\caption{The scenario to be evaluated in the simulation. An autonomous vehicle interacts with a crowd of crossing pedestrians. There is no road layout, so that the vehicle and pedestrians have the same priority. The objective of the autonomous vehicle is to keep a safe distance to the closest pedestrian in front while tries to keep the desired longitudinal speed as much as possible. It is assumed that the vehicle can only move longitudinally, so this study doesn't consider steering action.}
	\label{fig:scenario}
\end{figure}

A classical pedestrian crossing scenario was designed to evaluate the proposed MPC, as illustrated in figure~\ref{fig:scenario}. The actual (not predicted) pedestrian motion is also generated by aforementioned VCI model~\cite{yang2018social}. The simulation was repeatedly conducted for 2000 times. For each simulation, pedestrians were randomly initialized inside a rectangular area, so that situations of different pedestrian patterns can be covered. 

The major evaluation criteria is the time spent for the autonomous vehicle to complete the vehicle-pedestrian interaction. The same number of simulations with pure PID approach on the vehicle was also conducted for the comparison purpose, because PID is regarded as the most efficient traditional approach for the longitudinal speed regulation.  

Either using MPC or PID, different simulation might generate different interaction results, in which the vehicle might stop and wait for the pedestrian crossing, or directly drive through the pedestrian crowd without stopping and waiting. The reason is that when pedestrians interact with the autonomous vehicle, different pedestrian positions at any time $t=t'$ result in different vehicle speed regulation, which further increases the uncertainty of pedestrian motion at time $t>t'$. Therefore, the simulation was evaluated based on 3 different situations:
\begin{itemize}
    \item General Situation: consider the entire simulation results.
    \item Stop-and-Wait Situation: consider situations when both MPC and PID approaches stop and wait for pedestrian crossing.
    \item Non-stop Situation: consider situations when both MPC and PID approaches do not stop and wait.
\end{itemize}






\section{Result}
\subsection{Comparison Between MPC and PID}
\begin{figure*}
	\centering
	\includegraphics[width=\linewidth]{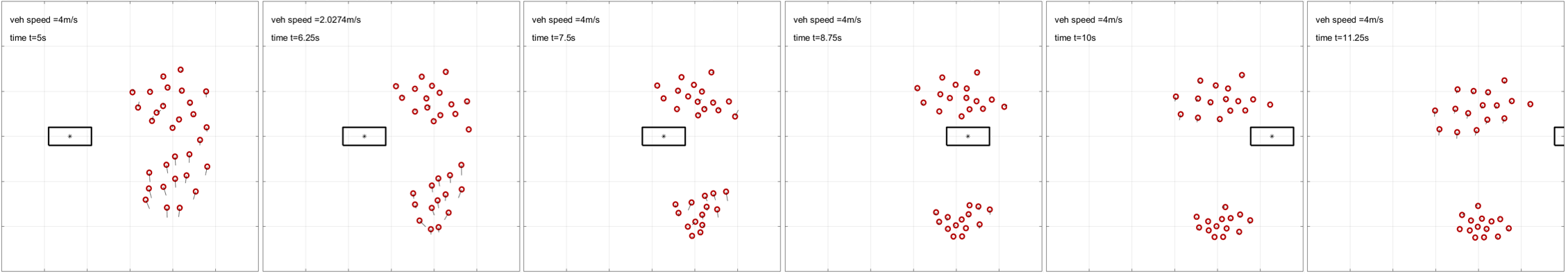}
	\caption{Screen-shots of an example of simulation at $t=5,6.25,7.5,8.75,10,11.25$ (s), respectively. Red circles indicate crossing pedestrians. Black rectangle indicates the autonomous vehicle using MPC approach.}
	\label{fig:sim_example_screenshots}
\end{figure*}

To visually illustrate the simulation result, figure~\ref{fig:sim_example_screenshots} shows the screen-shots of one simulation example. The corresponding video is available online.\footnote{\url{https://youtu.be/JlR3aZ1saDU}} In this example, the autonomous vehicle slightly adjusted its longitudinal speed and successfully completed the vehicle-pedestrian interaction.

\begin{figure*}
	\centering
	\includegraphics[width=\linewidth]{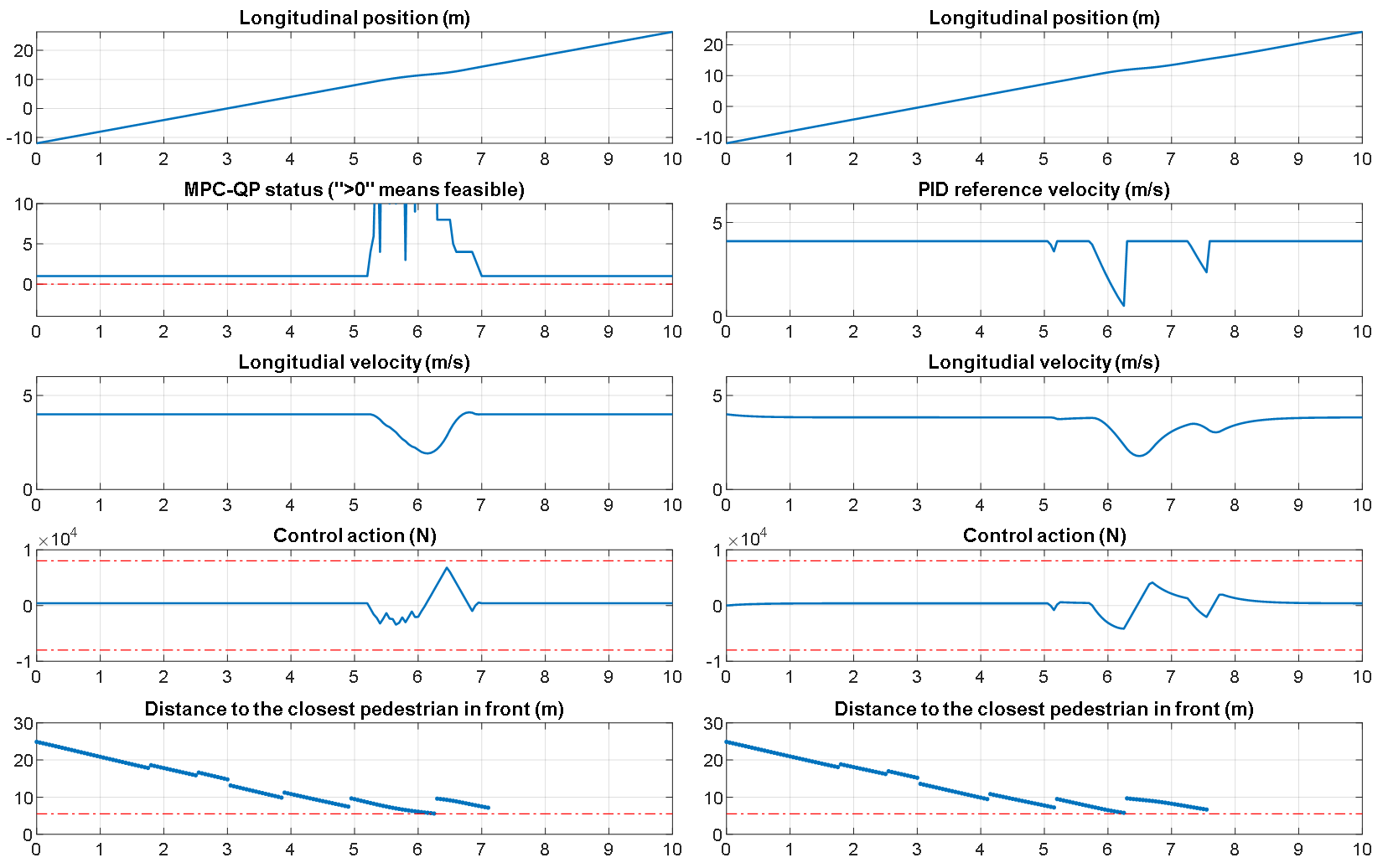}
	\caption{An example of performance comparison between MPC approach (left column) and PID approach (right column). In general, since MPC approach can predict the future trajectories of pedestrians, it generates smoother longitudinal velocity than PID approach. Note that half of the vehicle length (2.5m) is subtracted in the distance to the closest pedestrian in front.}
	\label{fig:sim_example_veh_state}
\end{figure*}

Figure~\ref{fig:sim_example_veh_state} shows the change of the vehicle state and controller state. In this particular example, MPC approach generates smoother longitudinal speed than PID approach.

The 2000 simulation results in scenarios of a number of 30 pedestrians were used for further analysis, which is divided into following 3 situations:
\begin{itemize}
    \item Difference of total time spent to complete the interaction in General Situation.
    \item Difference of longest time spent to wait for pedestrian crossing in Stop-and-Wait Situation.
    \item Difference of total time spent to complete the interaction in Non-stop Situation.
\end{itemize}

\begin{figure}
	\centering
	\includegraphics[width=\linewidth]{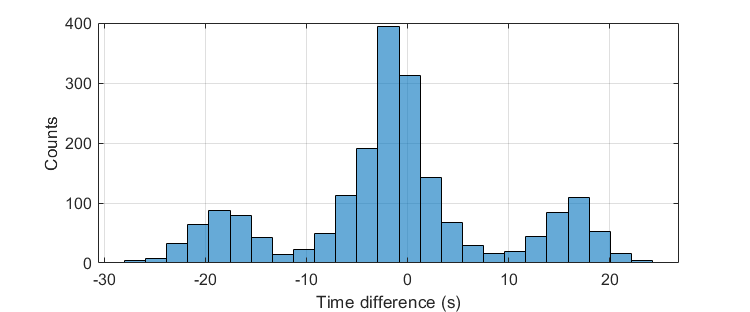}
	\caption{The difference of total time spent to complete the interaction between MPC approach and PID approach in all situations. The histogram is almost symmetric with a slight shift to the left (approximately 1s), which indicates the total time spent in MPC approach is generally shorter than PID approach.}
	\label{fig:time_difference_overall}
\end{figure}

\begin{figure}
	\centering
	\includegraphics[width=\linewidth]{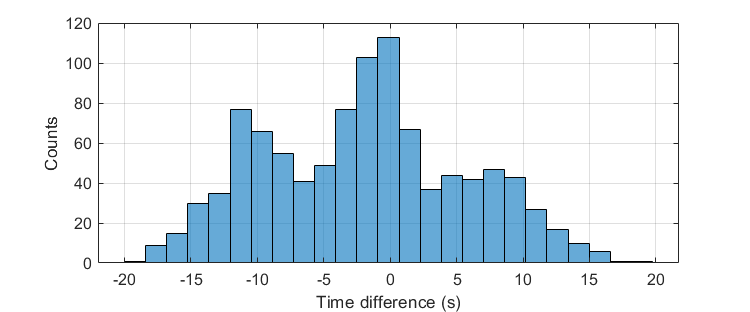}
	\caption{The difference of the longest time spent to wait for pedestrian crossing between MPC approach and PID approach in situations where both approaches stop and wait for pedestrian crossing. In the histogram, the slight shift to the left indicates the longest waiting time in MPC approach is generally shorter than PID approach.}
	\label{fig:time_difference_stop_longest}
\end{figure}

\begin{figure}
	\centering
	\includegraphics[width=\linewidth]{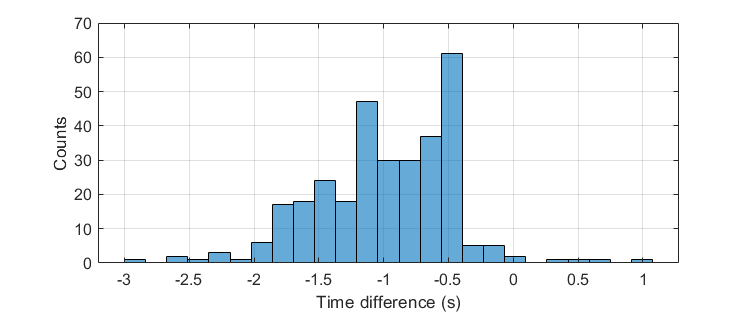}
	\caption{The difference of total time spent to complete the interaction between MPC approach and PID approach in situations where both approaches do not stop and wait. The histogram strongly indicates that the total time spent in MPC approach is shorter than PID approach.}
	\label{fig:time_difference_non_stop}
\end{figure}

Figures~\ref{fig:time_difference_overall}, \ref{fig:time_difference_stop_longest}, and \ref{fig:time_difference_non_stop} show the histograms of the time difference for the above 3 situations. In general, the MPC approach is better than PID approach. Detail description can be found in the figure captions.

\subsection{Different Pedestrian Density}
Simulations of different pedestrian density were also conducted. The numerical results are shown in table~\ref{tab:different_ped_density}.\footnote{N.A. in column 3 row 4: The number of instances in this situation is very small, hence the result is not provided here.} Generally speaking, the MPC approach is better than PID approach in terms of the time to complete the interaction, although the performance degrades as the pedestrian density decreases.

\begin{table}[b]
	\centering
	\caption{Average Time Difference (in seconds) Between MPC and PID Approaches with Different Pedestrian Density in Different Situations (A. General: time spent to complete the interaction in all situations; B. Stop-and-Wait: longest waiting time when both approaches stop and wait for pedestrian crossing; C. Non-stop: time spent to complete the interaction when both approaches do not stop and wait)}
	\label{tab:different_ped_density}
	\begin{tabular}{|c|c|c|c|}
		\hline
		\# of Ped. & General & Stop-and-Wait & Non-stop  \\ \hline
		30 &  -1.2665  &   -1.9457 & -0.9843 \\ \hline
		20  &  -0.5243  &  -1.8338 &  -0.7630\\ \hline
		10  &  -0.4153  &   N.A. & -0.5394\\ \hline
 
	\end{tabular}
\end{table}

There is a steady-state error of $\approx 0.16m/s$ at the desired speed $v_r=4m/s$ for the PID approach. The maximum delay caused by this steady-state error to complete the interaction is $\approx 0.4s$, which is calculated by assuming $v_r^{PID}=v_r$ all the time. Therefore, if the maximum delay of PID is considered, MPC approach is still better than PID approach in pedestrian-dense scenario (30 pedestrians in the simulation). However, in less-dense scenarios (20 or 10 pedestrians), it is hard to conclude that MPC approach is better than PID approach, although the simulation result still shows negative time difference.

\section{Conclusion}

This study investigated the possibility of applying model predictive control (MPC) supplemented with social force based vehicle-crowd interaction (VCI) model to regulate the longitudinal speed of the autonomous vehicle that faces a crowd of crossing pedestrians. The MPC problem was formulated based on state constraints and a safe distance to achieve collision avoidance and maximumly maintaining desired speed. The formulation was successfully converted into a standard quadratic programming (QP) problem, which can be easily solved by standard QP toolbox. Preliminary results demonstrated the merits of the proposed MPC approach by comparing it with classical pure PID approach.  

Future work is required to solve following issues:
\begin{itemize}
    \item In the pedestrian motion prediction process, this constant vehicle speed assumption can be improved by incorporating the VCI model into the MPC synthesis. Because of the non-linearity of VCI model, this incorporation requires modifying the VCI model so that the MPC can be properly synthesized and sucessfully solved.
    \item The performance of the PID approach can be improved by systematically tuning the PID parameters. Specifically, the steady-state error should be minimized or eliminated, and other effects such as rise time, overshot, settling time, and stability should also be carefully treated.
    \item In addition to hard constraints on the control action, a quadratic term of control effort could also be included in the MPC cost function, so that the overall MPC performance can be improved by taking the control action in consideration.
\end{itemize}

\bibliographystyle{ieeetr}
\bibliography{thebib}

\end{document}